\documentstyle[12pt]{article} \tolerance = 10000
\def\case#1/#2{\textstyle\frac{#1}{#2} }
\begin{document}
\title{Irrotational dust with $div ~~H = 0$}
\author{ {\sc W. M. Lesame$^{1,2}$,
G. F. R. Ellis$^{1}$, and P. K. S. Dunsby$^{1}$}\\
\normalsize{$1$ \it Department of Applied Mathematics,
University of Cape Town, }\\
\normalsize{ \it Rondebosch 7700, Cape Town, South Africa.} \\
\normalsize{$2$ \it Department of Applied Mathematics, University of Fort
Hare,} \\
\normalsize{\it Private Bag X1314, Alice 5700, South Africa.}}
\date{$\mbox{}$ \vspace*{0.3truecm} \\ \normalsize  \today}
\maketitle
\begin{abstract}
For irrotational dust the shear tensor is consistently diagonalizable
with its covariant time derivative: $\sigma_{ab}=0=\dot{\sigma}_{ab},\;
a\neq b$, if and only if the divergence of the magnetic part of the
Weyl tensor vanishes: $div~H = 0$.
We show here that in that case, the consistency of the Ricci constraints
requires that the magnetic part of the Weyl tensor itself
vanishes: $H_{ab}=0$.
\end{abstract}
\begin{center}{\it Subject headings:}\\
cosmology\,-\,galaxies: clustering,
formation\,-\,hydrodynamics\,-\,relativity\,-\,exact solutions
\end{center}
\section{Introduction}
The local non\,-\,linear dynamics of irrotational dust with a purely
``electric'' Weyl tensor has been investigated by Matarresse {\it et al.}
\cite{bi:MPS1,bi:MPS2,bi:MPS3}. These authors characterized such solutions as
`silent universes', because the time\,-\,evolution equations form a set of
ordinary differential equations. In \cite{bi:LDE1} it was shown that all
the integrability conditions for this case are satisfied and are consistent,
provided the initial constraint equations are satisfied. Then the development
along each world line is independent of those along neighbouring world lines;
once the evolution is underway, no information is exchanged between
neighbouring particles.

The dynamics of such universes have been studied mostly by employing
an orthonormal tetrad, which is a simultaneous eigentetrad of
the shear $\sigma_{ab}$ and the ``electric'' part of the Weyl tensor
$E_{ab}$. Variables defined in terms of these tetrads allow
development of useful phase\,-\,plane portraits \cite{bi:BMP}, which show
{\it inter alia} that the generic collapse configuration is a cigar or
spindle rather than a pancake.

In the linear theory, the magnetic part of the Weyl tensor $H_{ab}$ only
contains vector and tensor modes \cite{bi:BDE1}; also if vorticity is zero,
no vector modes are present, and $H_{ab}$ at most contains tensor modes
with vanishing divergence, which describe gravitational waves. In the
non\,-\,linear theory the physical content of $H_{ab}$ is however less clear.
At the core of the dynamics of silent universe is the assumption that in the
absence of $H_{ab}$, no gravitational waves occur.

For irrotational dust the shear tensor is consistently diagonalizable
with its covariant time derivative: $\sigma_{ab}=0=\dot{\sigma}_{ab},\;
a\neq b$, if and only if the divergence of the magnetic Weyl tensor
vanishes. Furthermore if the electric Weyl tensor vanishes then the
magnetic Weyl tensor vanishes \cite{bi:LES1}. So for irrotational dust
there are no solutions with a purely magnetic Weyl tensor.

The aim of this paper is to prove that for irrotational dust there are
no consistent solutions with a non-vanishing magnetic part of the Weyl tensor
that has a vanishing divergence. Thus the only consistent solutions
with vanishing divergence of $H_{ab}$ are those described by the silent
universes [2,7-9]. Mathematically, we start from the requirement that
the shear tensor $\sigma_{ab}$ is consistently diagonalizable with
$\dot{\sigma}_{ab}$. We then prove that this is possible iff $div H=0$
(\cite{bi:LES1}). This is a dynamical restriction, for it is a constraint
on the gravitational tidal field.
The implication of this is seen in the time derivative of the Ricci
constraints (which prescribes  spatial restrictions on both variables
$\sigma_{ab}, ~H_{ab}$), and  this yields $H_{ab}=0$.

{\it Notation}: Latin indices run from 0 to 3, and Greek indices from 1
to 3; semicolons denote covariant derivatives. Covariant differentiation
along the velocity vector $u^{a}$ is denoted by $(^{.})$ e.g., the
acceleration vector is $a_a \equiv \dot{u}_{a}:=u_{a;b}u^{b}$.
Kinematic and dynamic quantities are the same as in the other
papers of this series (they are comprehensively defined in
\cite{bi:ellis2}).
\section{Equations for Kinematic and Dynamic Quantities}   \label{sec-kine}
\subsection{Conservation Equations}
For a perfect fluid, the fluid acceleration is only determined by pressure
gradients so the restriction of vanishing pressure implies that
$a^a=0$. This means that each fluid element moves along a geodesic.
The conservation of energy and momentum $T^{ab}{}_{;b}=0$
leads to only one equation, the continuity equation:
\begin{equation}
\dot{\rho}=-\rho\Theta\;.
\label{eq:1.1}
\end{equation}
\subsection{The Ricci identities}
With the second restriction of vanishing vorticity $\omega_{ab}=0$,
the equations for the kinematic quantities follow from the Ricci identity:
$u_{a;d;c}-u_{a;c;d}=R_{abcd}u^b$.
\subsubsection{Propagation equations}
The expansion scalar $\Theta$ obeys
the Raychaudhuri equation:
\begin{equation}
\dot{\Theta}+\case{1}/{3}\Theta^2+2\sigma^2+\case{1}/{2}\kappa\rho=0\;,
\label{eq:1.2}
\end{equation}
where $\sigma^2\equiv\case{1}/{2}\sigma^{ab}\sigma_{ab}$ is the shear
scalar and $\kappa=8\pi G$ is the gravitational constant.
The remaining kinematic evolution equation is for the shear and is given by:
\begin{equation}
\dot{\sigma}_{ab}+\sigma_{ac}\sigma^c{}_b-\case{2}/{3}\sigma^2h_{ab}
+\case{2}/{3}\Theta\sigma_{ab}+E_{ab}=0\;,
\label{eq:1.4}
\end{equation}
where $E_{ac}=E_{(ac)}\equiv C_{abcd}u^bu^d$ is the ``electric'' part of
the Weyl tensor $C_{abcd}$ (satisfying $E_{ac}u^c=0$, $E^a{}_a = 0$).
$E_{ab}$ is that part of the gravitational field which describes tidal
interactions. The Weyl tensor can be decomposed into $E_{ab}$ and another
tensor called the ``magnetic'' part: $H_{ac} =
H_{(ac)} \equiv \case{1}/{2} \eta_{ab}{}^{gh} C_{ghcd}u^bu^d$ (satisfying
$H_{ac}u^c = 0$, $H^a{}_a=0$). This is the part of the gravitational field
that describes gravo\,-\,magnetic effects, and gravitational waves.
\subsubsection{Constraint equations}
Besides the evolution equations for the kinematical equations,
there are several constraints that our variables must satisfy.
On setting $p=\omega_{ab}=0$ we obtain as non\,-\,trivial constraints,
\begin{equation}
h^e{}_b\left(\case{2}/{3}\Theta^{;b}-h^d{}_c\sigma^{bc}{}_{;d}\right)=0\;,
\label{eq:1.5}
\end{equation}
the `$(0,\nu)$' field equations, and
\begin{equation}
H_{ad} = - h^t{}_ah^s{}_d\sigma_{(t}{}^{b;c}\eta_{s)fbc}u^{f} \;.
\label{eq:1.6}
\end{equation}
which  we refer to as the `$H_{ab}$' constraint, also (\ref{eq:1.5}) and
(\ref{eq:1.6}) are both referred to as Ricci constraints.
As we are considering the zero\,-\,vorticity case, we can also write down
the Gauss\,-\,Codacci equations for the 3\,-\,spaces orthogonal to $u^a$
(see \cite{bi:ellis2}); however we do not need them for what follows.

\subsection{Bianchi identities}
Additionally, the Bianchi identities must be satisfied, as they are the
integrability conditions for the other equations.

With the restrictions we have put on so far, they take the form:
\begin{equation}
h^t{}_ah^d{}_sE^{as}{}_{;d} - \eta^{tbpq}u_{b}\sigma^{d}{}_{q}H_{dq} =
\case{1}/{3}h^t{}_b\rho^{;b}\;,
\label{eq:1.7}
\end{equation}
\begin{equation}
h^m{}_ah^t{}_c\dot{E}^{ac}+h^{mt}\sigma^{ab}E_{ab}+\Theta E^{mt}
+ J^{mt} -3E_s{}^{(m}\sigma^{t)s}
=-\case{1}/{2}\rho\sigma^{tm}\;,
\label{eq:1.8}
\end{equation}
which we refer to as the ``div $E$'' and ``$\dot{E}$'' equations respectively,
and
\begin{equation}
h^{t}_{a}H^{as}{}_{;d}h^{d}_{s} + \eta^{tbpq}u_b \sigma^d{}_p E_{qd}= 0\;,
\label{eq:1.9}
\end{equation}
\begin{equation}
h^m{}_ah^t{}_c\dot{H}^{ac}+h^{mt}\sigma^{ab}H_{ab}+\Theta H^{mt}
- I^{mt} -3H_s{}^{(m}\sigma^{t)s}=0\;,
\label{eq:1.10}
\end{equation}
which are the ``div $H$'' and ``$\dot{H}$'' equations respectively,
where we have defined the curls of $E$ and $H$ respectively as
\begin{equation}
\mbox{'Curl E'}: \;\;\; I^{mt} = - h_a{}^{(m}\eta^{t)rsd}u_{r}E^{a}{}_{s;d}{}
\label{eq:1.11}
\end{equation}
\begin{equation}
\mbox{'Curl H'}: \;\;\; J^{mt} = h_a{}^{(m}\eta^{t)rsd}u_{r}H^{a}{}_{s;d}\;.
\label{eq:1.12a}
\end{equation}
The Bianchi identities are analogous to Maxwell's electromagnetic
equation \cite{bi:ellis2}. The gradient
$\case{1}/{3}h^{b}_{a}\rho_{,b}$ acts as a source
of the divergence of the $E_{ab}$ field in the ``div E'' constraint. For
zero vorticity the $H_{ab}$ field is source\,-\,free in the ``div H''
constraint. In the case of the time\,-\,development equations the ``curl E''
term $I^{mt}$ acts as a source of $\dot{H}$. On the other hand
$\case{1}/{2}\rho\sigma^{tm}$ acts as a source of $\dot{E}$,
as well as the ``curl H'' source term $J^{mt}$.

\section{Tetrad Description} \label{sec-tetrad}
We now show that the shear tensor $\sigma_{ab}$ is
diagonalizable in the same principal frame as $\dot{\sigma}_{ab}$
iff $div ~H_{ab} = 0$.

We first introduce an orthonormal tetrad that diagonalizes the shear
tensor i.e,
\begin{equation}
 \sigma_{ab} = 0 \;\;\; (a \neq b) \; .
\label{eq:tet1}
\end{equation}
This immediately
implies $\partial \sigma_{ab}/\partial \tau = 0$ for $(a \neq b)$, where
${\bf u} = \partial/\partial \tau$, but this gives no direct restrictions
on $\dot{\sigma}_{ab}$ because of the rotation coefficient terms in those
quantities. However from (\ref{eq:1.4}) the
tetrad satisfying (\ref{eq:tet1}) also satisfies
$\dot{\sigma}_{ab} = 0$ for $a\neq b$ provided
\begin{equation}
E_{ab} =0 \;\;(a\neq b)
\label{eq:tet5}
\end{equation}
in this tetrad. From the ``div H'' equation (\ref{eq:1.9}), equations
(\ref{eq:tet1}) and (\ref{eq:tet5}) imply that
\begin{equation}
div H = h^{t}{}_{a}H^{as}{}_{;d}h^{d}{}_{s} =0\,.
\label{eq:tet1a}
\end{equation}
Conversely if $div \,H =0$ then
$\sigma_{ab}$ and $E_{ab}$ are simultaneously diagonalizable from
(\ref{eq:1.9}), as shown in \cite{bi:BR}; furthermore then
$\dot{\sigma}_{ab} =0$ for $a\neq b$ from (\ref{eq:1.4}). Hence,
For irrotational dust the shear tensor is consistently diagonalizable
with its covariant time derivative: $\sigma_{ab}=0=\dot{\sigma}_{ab}$ for
$a\neq b$, if and only if the divergence of the magnetic Weyl tensor vanishes.
\\

We now adopt the assumption that $div \, H =0$,
as defined by equation (\ref{eq:tet1a});
and choose the preferred frame that diagonalizes $\sigma_{ab}$ and
$\dot{\sigma_{ab}}$. For  simplicity we denote
$\sigma_{aa}$ (no sum) by $\sigma_{a}$, and $E_{aa}$ by $E_{a}$. Both the
shear tensor $\sigma_{ab}$ and the 'electric' part of the Weyl tensor $E_{ab}$
satisfy the trace-free property, i.e.,
\begin{eqnarray}
E_{1} + E_{2} +E_{3}& =& 0 \; ,  \nonumber \\
\sigma_{1} + \sigma_{2} +\sigma_{3}& =& 0 \; .
\label{eq:tet7}
\end{eqnarray}
If we  now write out $0=\dot{\sigma}_{ab} = \sigma_{ab;c}u^{c}$ for
$(a \neq b)$, we get the following tetrad relation:
\begin{equation}
(\sigma_{a}-\sigma_{b})\Gamma^{a}{}_{0b} =0;\;\; a\neq b
\label{eq:tet2}
\end{equation}
where $u^{c} = \delta^{c}{}_{0}$ and $u^{a}u_{a}=-1$.
For an arbitrary shear tensor one may deduce from (\ref{eq:tet2}) that
\begin{equation}
\Gamma^{a}{}_{0b} =0 ~~(a \neq b) \label{eq:tet3}\; ,
\label{eq:tet2a}
\end{equation}
and hence the tetrad is Fermi-propagated along
$u_{a}$. Condition (\ref{eq:tet2a}) can also be shown to be valid for the
case of degenerate shear. This is achieved by performing a rotation in the
degenerate plane and using this tetrad freedom to obtain (\ref{eq:tet2a}),
see [6]. Furthermore for dust the vanishing of vorticity in a shear
eigenframe is equivalent to the conditions
\begin{equation}
\Gamma^{0}{}_{ab} =0\; .
\label{eq:tet4}
\end{equation}
Now by (\ref{eq:tet3}) and the diagonality of $E_{ab}$, it follows that
also $\dot{E}_{ab}$ is diagonal:
$$\dot{E}_{ab}= (E_a - E_b) \Gamma^a{}_{0b} = 0, ~~~a \neq b.$$
Then from the time development equation
(\ref{eq:1.8}) it follows that ``curl H'' is also diagonal i.e.,
\begin{equation}
J^{mt} = h_a{}^{(m}\eta^{t)rsd}u_{r}H^{a}{}_{s;d}=0\;, \;\;\;
m \neq t\;.
\label{eq:tet6}
\end{equation}
Consistent with the rest of the notation, we denote $J_{aa}$ by $J_{a}$.
Then the fact it is trace free becomes $J_{1} + J_{2} + J_{3}=0$.

\subsection{Propagation equations}
Equations (\ref{eq:1.1})-(\ref{eq:1.10}) governing the
evolution of irrotational dust may now be written
as a set of propagation equations tied to a set of constraint equations.

A direct conversion of  equations (\ref{eq:1.1}), (\ref{eq:1.2}),
(\ref{eq:1.4}) and (\ref{eq:1.8}) into a tetrad system which is an
eigenframe for both the shear tensor and the electric Weyl tensor yields
the following time-evolution equations:
\begin{eqnarray}
\dot{\rho} &=& - \rho \Theta\;, \nonumber\\
\dot{\Theta}&=&- \case{1}/{3}\Theta^{2}
- (\sigma_{1}{}^{2} + \sigma_{2}{}^{2} + \sigma_{3}{}^{2})
-\case{1}/{2}\rho\;, \nonumber\\
\dot{\sigma}_{\mu}&=&-(\sigma_{\mu})^{2} -\case{2}/{3}\Theta\sigma_{\mu}
+\case{1}/{3}(\sigma_{1}{}^{2} + \sigma_{2}{}^{2} + \sigma_{3}{}^{2})
- E_{\mu}\;, \label{eq:in3}   \nonumber\\
\dot{E}_{\mu}&=&-\Theta E_{\mu} -\case{1}/{2} \rho \sigma_{\mu}- J_{\mu}
+ 3\sigma_{\mu}E_{\mu} - (\sigma_{1}E_{1} + \sigma_{2}E_{2}
+ \sigma_{3}E_{3})\;,
\label{eq:prop1}
\end{eqnarray}
where $\mu = 1,2,3$ and no sum is carried out. The propagation equation
for the magnetic Weyl tensor $H_{ab}$ follows from
the Bianchi identity (\ref{eq:1.10}) and has the form
\begin{eqnarray}
\dot{H}_{11} &=& -\Theta H_{11} + 3 \sigma_{1}H_{11} + I_{11}
               -(\sigma_{1}H_{11} + \sigma_{2}H_{22} + \sigma_{3}H_{33})
\; , \nonumber \\
\dot{H}_{22} &=& -\Theta H_{22} + 3 \sigma_{2}H_{22} + I_{22}
               -(\sigma_{1}H_{11} + \sigma_{2}H_{22} + \sigma_{3}H_{33})
\; , \nonumber \\
\dot{H}_{33} &=& -\Theta H_{33} + 3 \sigma_{3}H_{33} + I_{33}
               -(\sigma_{1}H_{11} + \sigma_{2}H_{22} + \sigma_{3}H_{33})
\; , \label{eq:prop2}    \\
\dot{H}_{23} &=& -\Theta H_{23} - \case{3}/{2} \sigma_{1}H_{23} + I_{23}
\; , \nonumber \\
\dot{H}_{12} &=& -\Theta H_{12} - \case{3}/{2} \sigma_{3}H_{12} + I_{12}
\; , \nonumber \\
\dot{H}_{31} &=& -\Theta H_{31} - \case{3}/{2} \sigma_{2}H_{31} + I_{31}
\; . \label{eq:prop3}
\end{eqnarray}

It is important to note that in each of these equations, the left hand
side is in terms of the covariant derivative $\dot{T} = T_{;a} u^a$, and so
in general involves rotation coefficients; for example, in the obvious
notation, the left-hand side of the 3rd equation (\ref{eq:prop1}) is
\begin{equation}
\dot{\sigma}_\nu = \sigma_{\nu,0} - 2 \sigma_\nu \Gamma^\nu{}_{0\nu}~~(no~sum)
\label{eq:time1}
\end{equation}
and that of the 4th equation is
\begin{equation}
\dot{E}_\nu = E_{\nu,0} - 2 E_\nu \Gamma^\nu{}_{0\nu} ~~(no ~sum).
\label{eq:time2}
\end{equation}

\subsection{Ricci Constraints}
The tetrad form of the  Ricci constraints (\ref{eq:1.5}),
(\ref{eq:1.6}) are as follows: the ``$(0,\nu)$'' field equations
(\ref{eq:1.5}) take the form
\begin{eqnarray}
\case{2}/{3}\partial_{1}\Theta&=&\partial_{1}\sigma_{1} +
(\sigma_{1}-\sigma_{2})\Gamma^{2}{}_{21} +
(\sigma_{1}-\sigma_{3})\Gamma^{3}{}_{31}\;,
\nonumber  \\
\case{2}/{3}\partial_{2}\Theta&=&\partial_{2}\sigma_{2}
+ (\sigma_{2}-\sigma_{1})\Gamma^{1}{}_{12}
+ (\sigma_{2}-\sigma_{3})\Gamma^{3}{}_{32}\;,
\nonumber  \\
\case{2}/{3}\partial_{3}\Theta&=&\partial_{3}\sigma_{3}
+ (\sigma_{3}-\sigma_{1})\Gamma^{1}{}_{13}
+ (\sigma_{3}-\sigma_{2})\Gamma^{2}{}_{23}\;.
\label{eq:con1}
\end{eqnarray}
The ``$H_{ab}$'' equations (\ref{eq:1.6})
take the form
\begin{eqnarray}
H_{11} &=&
\Gamma^{1}{}_{23}(\sigma_{3}-\sigma_{1})
- \Gamma^{1}{}_{32}(\sigma_{2}-\sigma_{1})\;,       \nonumber  \\
H_{22} &=&
\Gamma^{2}{}_{31}(\sigma_{1}-\sigma_{2})-
\Gamma^{2}{}_{13}(\sigma_{3}-\sigma_{2})\;,         \nonumber  \\
H_{33} &=&
\Gamma^{3}{}_{12}(\sigma_{2}-\sigma_{3}) -
\Gamma^{3}{}_{21}(\sigma_{1}-\sigma_{3})\; ;
\label{eq:con2}
\end{eqnarray}
\begin{eqnarray}
H_{23} &=&
\case{1}/{2}\left[\partial_{1}(\sigma_{2}-\sigma_{3})
+ \Gamma^{3}{}_{31}(\sigma_{1}- \sigma_{3})
- \Gamma^{2}{}_{21}(\sigma_{1} -\sigma_{2})\right]\;, \nonumber \\
H_{31} &=&
\case{1}/{2}\left[\partial_{2}(\sigma_{3}-\sigma_{1})
+\Gamma^{3}{}_{32}(\sigma_{3}- \sigma_{2})
-\Gamma^{1}{}_{12}(\sigma_{1} -\sigma_{2})\right]\;,  \nonumber \\
H_{12} &=&
\case{1}/{2}\left[\partial_{3}(\sigma_{1}-\sigma_{2})
+\Gamma^{1}{}_{13}(\sigma_{1} - \sigma_{3})
-\Gamma^{2}{}_{23}(\sigma_{2} - \sigma_{3})\right]\;.
\label{eq:con3}
\end{eqnarray}

\subsection{Form of $H_{ab}$ }
Given our choice of a tetrad frame which diagonalizes $\sigma_{ab}$
(and hence according to theorem~1, $div H=0$) we show below that consistency
of the Ricci constraint requires that $H_{ab}=0$.

To calculate time derivatives of the constraints
we first start from the covariant form of the constraint equation,
say ($G$) and then take the time derivative as defined by
$(G\dot{)} \equiv (G)_{;a}u^{a}$. We commute derivatives and use the
constraint equations; the result is then converted to tetrad form.
We give an example of such a calculation in Appendix A (using the
'$H_{ab}$' constraint). This approach avoids any direct calculations of the
time derivatives of Ricci coefficients $\Gamma^{a}{}_{bc}$.

The time derivative of the  ``$(0,\nu)$'' constraint (\ref{eq:1.5}) in
covariant form is
\begin{equation}
0 = h^{eb} \left[ \case{2}/{3} \left( \Theta_{;b} \right)^{.}
            - h^{d}{}_{c} \left( \sigma_{b}{}^{c}{}_{;d} \right)^{.}
\right] \; .
\label{eq:der1.1}
\end{equation}
If we use
\begin{eqnarray}
\left( \Theta_{;b} \right)^{.} &=&  \left( \dot{\Theta} \right)_{;b}
                              -\Theta_{;p}u^{p}{}_{;b}
  \; ,  \\
\left( \sigma_{b}{}^{c}{}_{;d} \right)^{.} &=&
                \left( \dot{\sigma}_{b}{}^{c} \right)_{;d}
                      -\sigma_{b}{}^{c}{}_{;p} u^{p}{}_{;d}
               +R^{c}{}_{qpd}\sigma_{b}{}^{q}u^{p}
               -R^{q}{}_{bpd}\sigma_{q}{}^{c}u^{p}
\label{eq:der1.1a}
\end{eqnarray}
and substituting into (\ref{eq:der1.1}) we get
\begin{eqnarray}
0 &=& h^{eb} \left[
        \left(\dot{\Theta}\right)_{;b} -\Theta_{;p}u^{p}{}_{;b}
 -h^{d}{}_{c} \left\{ \left(\dot{\sigma}_{b}{}^{c}\right)_{;d}
                      -\sigma_{b}{}^{c}{}_{;p} u^{p}{}_{;d}
              \right\}  \right]
   \nonumber \\
 & & -h^{eb} h^{d}{}_{c}  \left\{ R^{c}{}_{qpd}\sigma_{b}{}^{q}u^{p}
               -R^{q}{}_{bpd}\sigma_{q}{}^{c}u^{p}
               \right\}
\; . \label{eq:der1.2}
\end{eqnarray}
We may now convert (\ref{eq:der1.2}) into tetrad form and use
(\ref{eq:prop1}) and (\ref{eq:con1}).
The Riemann tensor is given in terms of the Weyl tensor in Appendix~A.
On further simplification (see that Appendix for details) we obtain
\begin{eqnarray}
0=\left(\sigma_{2} -\sigma_{3}\right) H_{23}\; , ~~~
0=\left(\sigma_{3} -\sigma_{1}\right) H_{31}\; , ~~~
0=\left(\sigma_{1} -\sigma_{2}\right) H_{12}\; .
\label{eq:der1.3}
\end{eqnarray}

The time derivative of the  ``$H_{ab}$'' constraint
(\ref{eq:1.5}) in covariant form may be written as
\begin{equation}
\dot{H}_{ad} = h^{(t}{}_{a}H^{s)}{}_{d} \eta^{c}{}_{fbs}u^{f} \left[
          \left(\dot{\sigma}^{b}{}_{t}\right)_{;c}
          - \sigma^{b}{}_{t;p}u^{p}{}_{;c}
          + R^{b}{}_{qpc}\sigma^{q}{}_{t}u^{p}
          - R^{q}{}_{tpc}\sigma^{b}{}_{q}u^{p}
          \right] \; .
\label{eq:der1.4}
\end{equation}
The tetrad form of (\ref{eq:der1.4}) for $a\neq b$ yields
\begin{eqnarray}
0=\sigma_{1} H_{23}\; , ~~~0=\sigma_{2} H_{31}\; , ~~~0=\sigma_{3} H_{12}\;
\label{eq:der1.5}
\end{eqnarray}
(see the Appendix B for details).\\

For nonzero shear conditions (\ref{eq:der1.3}) and (\ref{eq:der1.5}) yields
\begin{equation}
H_{12}=H_{23}=H_{31} = 0,
\label{eq:der1.6}
\end{equation}
and thus the ``magnetic'' part of the Weyl tensor $H_{ab}$ is also diagonal.
Through the $\dot{H}$ equations, this implies $curl ~E$ too is diagonal.
We henceforth write $H_{aa}$ as $H_{a}$; then
the trace free property of
$H_{ab}$ is
\begin{equation}
H_{1} + H_{2} +H_{3} =0 \; .
\label{eq:der1.7}
\end{equation}
The tetrad form of (\ref{eq:der1.4}) for $a=b$ yields (see Appendix B)
\begin{eqnarray}
\sigma_{1}H_{1}= \sigma_{2}H_{2}=\sigma_{3}H_{3}\; ,
\label{eq:der1.8}
\end{eqnarray}
We point out that diagonal property (\ref{eq:der1.6}) of $H_{ab}$ was not used
in obtaining (\ref{eq:der1.8}).
We now introduce a constant $\lambda$ that relates the shear eigenvalue
$\sigma_{1}= \sigma$ to $\sigma_{2}$ as follows
\begin{equation}
\sigma_{1} = \sigma,\;\;\; \sigma_{2}= \lambda \sigma \; .
\label{eq:der1.10}
\end{equation}
If the shear tensor is degenerate in the $e_{1},e_{2}$ plane then $\lambda =
1$.
Now equation (\ref{eq:der1.8}) $\sigma_{1}H_{1}= \sigma_{2}H_{2}$
prescribes the following relation on the eigenvalues $H_{1}, H_{2}$
of the magnetic Weyl tensor:
\begin{equation}
H_{1}= \lambda H, \;\;\; H_{2} = H \; .
\label{eq:der1.11}
\end{equation}
The trace-free property yields
\begin{equation}
\sigma_{3} = -(1+\lambda)\sigma\,,\;\;\;
H_{3}= -(1+\lambda) H \; ,
\label{eq:der1.12}
\end{equation}
so now from (\ref{eq:der1.8}) if we use $\sigma_{1}H_{1} = \sigma_{3}H_{3}$
we obtain
\begin{equation}
0 = (1+\lambda + \lambda^{2}) \sigma H
\label{eq:der1.13}
\end{equation}
The following cases satisfy (\ref{eq:der1.13})

{\bf Case} A: $H=0$ with $\sigma \neq 0$ and  $\lambda$ remaining arbitrary.
This case include degenerate shear, $\lambda =1$, and is studied in
\cite{bi:BMP, bi:MPS1, bi:MPS2, bi:MPS3} and \cite{bi:LDE1}. Variables defined
in terms of the above tetrad frame allow
development of useful phase\,-\,plane portraits, which show
{\it inter alia} that the generic collapse configuration is a cigar or
spindle rather than a pancake.

{\bf Case} B: $\sigma =0$. For this case all the shear eigenvalues vanish and
this implies $H=0$ and $E=0$ and the model is FRW.

{\bf Case} C: $(1+\lambda + \lambda^{2}) =0$. The values of $\lambda$ are
complex. The magnitudes of the shear tensor $\sigma_{mag}$ and the
magnetic Weyl tensor $H_{mag}$
have the tetrad form
\begin{eqnarray}
\sigma_{mag}=
\case{1}/{2}(\sigma_{1}{}^{2} + \sigma_{2}{}^{2} + \sigma_{3}{}^{2}).
\nonumber \\
H_{mag} = H_{1}{}^{2} + H_{2}{}^{2} + H_{3}{}^{2} \; .
\end{eqnarray}
On using (\ref{eq:der1.10}) and (\ref{eq:der1.12}) for this case we get
\begin{equation}
\sigma_{mag} =(1+\lambda + \lambda^{2})\sigma^{2} = 0,
{}~~~\nonumber
H_{mag} = (1+\lambda + \lambda^{2}) H^{2} = 0
\end{equation}
and hence both shear and the magnetic Weyl tensor are zero, leading to the
same results as case B above.

Thus we can formulate the following
\begin{quote}
{\bf Theorem:}\\
For irrotational dust the divergence of the magnetic Weyl tensor
vanishes, $div H=0$ (or equivalently the shear tensor
is consistently diagonalizable) if and only if the magnetic
Weyl tensor vanishes, $H_{ab}=0$.
\end{quote}

\section{Conclusion}
For irrotational dust $\rho\neq 0$ the existence of a tetrad frame which
is a principal frame for the shear tensor $\sigma_{ab}$ and its covariant time
derivative $\dot{\sigma}_{ab}$ requires that the divergence of
the magnetic Weyl tensor vanish. We have shown here that if we employ
this frame then the magnetic Weyl tensor itself vanishes; $H_{ab}=0$, as a
consistency requirement of the Ricci constraints. This establishes a new
property, that the only consistent solutions for irrotational dust
with vanishing $div~H$ are those described as the `silent' universe
\cite{bi:BMP},\cite{bi:LDE1}. Hence gravitational waves interacting with
irrotational dust will have to have $div ~H \neq 0$, contrary to the usual
result of linearised theory.

The key point here is that (assuming the fluid is irrotational), our result
comes from the second term in the `div H' equation (8); but when we
linearise, that term is necessarily second order.
Thus always $Div H=0$ (to first order) in the linear case.
Hence our exact result comes from a term which plays no role in the
linearised theory, if we discard all second order terms in all equations;
in this case the surviving Riemann terms would be disregarded (as they
always consist of a Weyl tensor component, which is first order,
multiplying a first order quantity: see (32), (34), (37)).

However this argument needs to be treated with care. One needs to recall that
second order terms can only be dropped from an equation if there is a
non-zero first order term present, so that the second-order term is negligible
relative to the first-order one. However this argument cannot be applied to
the key equation (8): both terms are second order, and as there is no first
order term, we have to ensure that the equation is true to second order
accuracy. Thus even in linearised theory, we cannot ignore this second-order
equation.
The following intriguing situation results: if we have a linearised
solution where $H_{ab} \neq 0$ and $div ~H$ is non-zero but second order,
we can presumably get a consistent solution. If however $H_{ab} \neq 0$ with
$div~H$ {\it exactly} zero, the solution will not be consistent - because of
the above proof.

Thus in the linear theory, it is possible to have models where $Div~H=0$ to
second order but $H_{ab}\neq 0$, and indeed that is usually assumed for
gravitational waves. Our result (for irrotational dust only) shows that in
this case in fact we must
have $div~H \neq 0$, although it is second order.
We have therefore an example of linearization instability
in that the usual process of linearization leads to a different answer than
the exact result - which in fact constrains the linearised solution, even
though this is usually not commented on.

\section*{Acknowledgements}
We thank the FRD (South Africa) for financial support, and a referee for
catching an important error in an earlier version of this paper (which
claimed a different result).


\appendix
\section{Time propagation of the ``$(0,\mu)$'' constraint}

The covariant form of the ``$(0,\mu)$'' constraint (4) is
\begin{equation}
h^e{}_b\left(\case{2}/{3}\Theta^{;b}-h^d{}_c\sigma^{bc}{}_{;d}\right)=0\;,
\label{apb:b1}
\end{equation}
with tetrad form
\begin{eqnarray}
\case{2}/{3}\partial_{1}\Theta&=&\partial_{1}\sigma_{1} +
(\sigma_{1}-\sigma_{2})\Gamma^{2}{}_{21} +
(\sigma_{1}-\sigma_{3})\Gamma^{3}{}_{31}\;,
\nonumber  \\
\case{2}/{3}\partial_{2}\Theta&=&\partial_{2}\sigma_{2}
+ (\sigma_{2}-\sigma_{1})\Gamma^{1}{}_{12}
(\sigma_{2}-\sigma_{3})\Gamma^{3}{}_{32}\;,
\nonumber  \\
\case{2}/{3}\partial_{3}\Theta&=&\partial_{3}\sigma_{3}
+ (\sigma_{3}-\sigma_{1})\Gamma^{1}{}_{13}
+ (\sigma_{3}-\sigma_{2})\Gamma^{2}{}_{23}\;,
\label{apb:b2}
\end{eqnarray}
The covariant time propagation of (\ref{apb:b1}) is (31),
\begin{eqnarray}
0 &=& h^{eb} \left[ \case{2}/{3} \left( \Theta_{;b} \right)^{.}
            - h^{d}{}_{c} \left( \sigma_{b}{}^{c}{}_{;d} \right)^{.}
                    \right]
\nonumber  \\
 &=& h^{eb} \left[
        \left(\dot{\Theta}\right)_{;b} -\Theta_{;p}u^{p}{}_{;b}
 -h^{d}{}_{c} \left\{ \left(\dot{\sigma}_{b}{}^{c}\right)_{;d}
                      -\sigma_{b}{}^{c}{}_{;p} u^{p}{}_{;d}
              \right\}  \right]
   \nonumber \\
 & & -h^{eb} h^{d}{}_{c}  \left\{ R^{c}{}_{qpd}\sigma_{b}{}^{q}u^{p}
               -R^{q}{}_{bpd}\sigma_{q}{}^{c}u^{p}
               \right\} \; .
\label{apb:b3}
\end{eqnarray}
We now  write the Riemann tensor $R^{s}{}_{mpc}$  in the
terms of the Weyl tensor as
\begin{eqnarray}
R^{s}{}_{mpc}&=& C^{s}{}_{mpc} + \case{1}/{2}(g^{s}{}_{p}R_{cm}
        + g^{s}{}_{c}R_{pm} - g_{mp}R^{s}{}_{c} + g_{mc}R^{s}{}_{p})
  \nonumber \\
     &\mbox{}& - \frac{R}{6}( g^{s}{}_{p}g_{cm}-g^{s}{}_{c}g_{pm})
\label{apb:b5}
\end{eqnarray}
where
\begin{eqnarray}
C^{s}{}_{mpc} &\equiv&  (\eta^{s}{}_{mij}\eta_{pckl} +
g^{s}{}_{mij}g_{pckl})u^{i}u^{k}E^{bd}
   \;  \nonumber \\
   & &   + (\eta^{s}{}_{mij} g_{pckl} +
   \eta_{pckl}g^{s}{}_{mij})u^{i}u^{k}H^{bd}
   \; ;
\label{apb:b6}
\end{eqnarray}
and
\begin{equation}
g^{s}{}_{mij} \equiv g^{s}{}_{i}g_{mj} - g^{s}{}_{j}g_{mi}\; .
\label{apb:b7}
\end{equation}
The two terms involving the Riemann tensor $R^{a}{}_{bcd}$
in (\ref{apb:b3}) may be written in covariant form as
\begin{eqnarray}
(RT)^{e} &= & -h^{eb} h^{d}{}_{c}  \left[ R^{c}{}_{qpd}\sigma_{b}{}^{q}u^{p}
               -R^{q}{}_{bpd}\sigma_{q}{}^{c}u^{p} \right]
\nonumber \\
\nonumber \\
 &=& -h^{eb} h^{d}{}_{c} u^{p} \left[
 \case{1}/{2}\sigma^{q}{}_{b}
     \left(g^{c}{}_{p}R_{qd} + g^{c}{}_{d}R_{qp}
                -g_{qp}R^{c}{}_{d} + g_{qd}R^{c}{}_{p} \right) \right.
\nonumber \\
    &     &
 - \case{1}/{2}\sigma^{c}{}_{q}
     \left( g^{q}{}_{p}R_{bd} + g^{q}{}_{d}R_{bp}
                -g_{bp}R^{q}{}_{d} + g_{bd}R^{q}{}_{p} \right)
\nonumber \\
       &  &  \left.
- \case{R}/{6} \sigma^{q}{}_{b}
    \left(g^{c}{}_{p}g_{qd} - g^{c}{}_{d}g_{qp}\right)
+ \case{R}/{6} \sigma^{c}{}_{q}
    \left(g^{q}{}_{p}g_{bd} - g^{q}{}_{d}g_{bp}\right) \right]
\nonumber   \\
 & &  \nonumber\\
& - & h^{eb} h^{d}{}_{c} u^{p} u^{i}u^{k} E^{jl}\left[
\sigma^{q}{}_{b} \eta^{c}{}_{qij}\eta_{pdkl}
+\sigma^{q}{}_{b}\left(g^{c}{}_{i}g_{qj} - g^{c}{}_{j}g_{qi}\right)
    \left(g_{pk}g_{dl} - g_{pl}g_{dk}\right)  \right.
\nonumber  \\
       &  &  \left.
-\sigma^{c}{}_{q}\eta^{q}{}_{bij}\eta_{pdkl}
-\sigma^{c}{}_{q} \left(g^{q}{}_{i}g_{bj} - g^{q}{}_{j}g_{bi} \right)
    (g_{pk}g_{dl} - g_{pl}g_{dk})
               \right]
\nonumber   \\
 & &  \nonumber\\
&- & h^{eb} h^{d}{}_{c} u^{p} u^{i}u^{k} H^{jl}\left[
\sigma^{q}{}_{b} \eta^{c}{}_{qij} \left(g_{pk}g_{dl} - g_{pl}g_{dk}\right)
-\sigma^{c}{}_{q} \eta^{q}{}_{bij} \left(g_{pk}g_{dl} - g_{pl}g_{dk}\right)
  \right.
\nonumber  \\
      &   & \left.
+ \sigma^{q}{}_{b}\eta_{pdkl}\left(g^{c}{}_{i}g_{qj} - g^{c}{}_{j}g_{qi}\right)
-\sigma^{c}{}_{q} \eta_{pdkl}\left(g^{q}{}_{i}g_{cj} - g^{q}{}_{j}g_{ci}\right)
                   \right]
\; .
\label{apb:b7a}
\end{eqnarray}
The only non-vanishing terms in (\ref{apb:b7a}) are
\begin{eqnarray}
(RT)^{e} &=&
-  h^{eb} h^{d}{}_{c} u^{p} u^{i}u^{k} H^{jl}\left[
\sigma^{q}{}_{b} \eta^{c}{}_{qij} g_{pk}g_{dl}
-\sigma^{c}{}_{q} \eta^{q}{}_{bij} g_{pk}g_{dl} \right]
\nonumber  \\
&= & \eta^{c}{}_{qij} \sigma^{eq}H^{j}{}_{c} u^{i}
-h^{eb} \eta^{q}{}_{bij}\sigma_{ql} H^{jl}u^{i}
\; .\label{apb:b8}
\end{eqnarray}
The tetrad form of (\ref{apb:b8}) becomes
\begin{eqnarray}
(RT)^{e} = \eta^{c}{}_{q0j} \sigma^{eq}H^{j}{}_{c}
- \eta^{q}{}_{e0j}\sigma_{ql} H^{jl}
\; .\label{apb:b9}
\end{eqnarray}
and from (\ref{apb:b9}) for $e=1$ we obtain
\begin{eqnarray}
(RT)^{1}
& = &\eta^{c}{}_{q0j} \sigma^{1q}H^{j}{}_{c}
         - \eta^{q}{}_{10j}\sigma_{ql} H^{jl}
\nonumber  \\
& = & \eta^{3}{}_{102} \sigma^{1}H^{2}{}_{3}
     + \eta^{2}{}_{103} \sigma^{1}H^{3}{}_{2}
         - \eta^{3}{}_{102}\sigma_{3} H^{23}
         - \eta^{2}{}_{103}\sigma_{2} H^{32}
\nonumber  \\
&=& H_{23}(\sigma_{3}-\sigma_{2})
\; .\label{apb:b10}
\end{eqnarray}
Similarly from (\ref{apb:b9}) for $e=2,3$ we get
\begin{eqnarray}
(RT)^{2} &=& H_{31}(\sigma_{1}-\sigma_{3}) \;,
\nonumber  \\
(RT)^{3} &=& H_{12}(\sigma_{2}-\sigma_{1}) \;.
\label{apb:b11}
\end{eqnarray}

We now focus on the tetrad form of the time propagation equation
(\ref{apb:b3}) for \underline{$e=1$}; in expanded form this becomes
\begin{eqnarray}
0 &=& \case{2}/{3} \left[ \partial_{1}\dot{\theta}
                  -\theta_{1}\partial_{1}\theta  \right]
-\left[ \partial_{1}\dot{\sigma_{1}}
             -\theta_{1}\partial_{1}\sigma_{1}  \right]
\nonumber \\
 && -\Gamma^{2}{}_{21}\left[( \dot{\sigma}_{1}-\dot{\sigma}_{2})
               -\theta_{2}(\sigma_{1}-\sigma_{2})  \right]
-\Gamma^{3}{}_{31}\left[( \dot{\sigma}_{1}-\dot{\sigma}_{3})
               -\theta_{3}(\sigma_{1}-\sigma_{3})  \right]
\nonumber  \\
&& + H_{23}(\sigma_{3}-\sigma_{2})
\; ,
\label{apb:b12}
\end{eqnarray}
where the last term on the right hand side is the contribution of the
Riemann term $(RT)^{1}$ calculated in (\ref{apb:b10}).
On using $\dot{\theta}$ and $\dot{\sigma}_{\mu}$ from equation (20)
we get
\begin{eqnarray}
0 &=& \case{2}/{3} \left[
  -\case{2}/{3}\theta\partial_{1}{\theta}
  -\partial_{1}(\sigma_{1}{}^{2}+\sigma_{2}{}^{2}+\sigma_{3}{}^{2})
  -\case{1}/{2}\partial_{1}\rho
  -(\sigma_{1}+\case{1}/{3}\theta)\partial_{1}\theta
  \right]
\nonumber  \\
&& -\left[ -2\sigma_{1}\partial_{1}\sigma_{1}
-\case{2}/{3}\theta\partial_{1}\sigma_{1}
-\case{2}/{3}\sigma_{1}\partial_{1}\theta
+\case{1}/{3}\partial_{1}(\sigma_{1}{}^{2}+\sigma_{2}{}^{2}+\sigma_{3}{}^{2})
-\sigma_{1}\partial_{1}\sigma_{1}   \right.
\nonumber  \\
&& \left.
-\partial_{1}E_{1}
-(\sigma_{1}+\case{1}/{3}\theta)\partial_{1}\sigma_{1}  \right]
\nonumber \\
 && -\Gamma^{2}{}_{21}\left[- (\sigma_{1}{}^{2}- \sigma_{2}{}^{2})
-\case{2}/{3}\theta(\sigma_{1}-\sigma_{2})
-(E_{1}-E_{2})
-(\sigma_{2}+\case{1}/{3}\theta)(\sigma_{1}-\sigma_{2})  \right]
\nonumber \\
 && -\Gamma^{3}{}_{31}\left[- (\sigma_{1}{}^{2}- \sigma_{3}{}^{2})
-\case{2}/{3}\theta(\sigma_{1}-\sigma_{3})
-(E_{1}-E_{3})
-(\sigma_{3}+\case{1}/{3}\theta)(\sigma_{1}-\sigma_{3})  \right]
\nonumber \\
&& + H_{23}(\sigma_{3}-\sigma_{2})
\; ,
\label{apb:b13}
\end{eqnarray}
where we also used $u_{a;b}= \sigma_{ab} +\case{1}/{3}\theta h_{ab}$ and
$\theta_{\mu} = \sigma_{\mu} + \case{1}/{3}\theta$.
We now apply to (\ref{apb:b13}) the ``$(0,\mu)$'' constraint (\ref{apb:b2})
and the ``$div~E$'' constraint written here
in tetrad form as
\begin{eqnarray}
\case{1}/{3}\partial_{1}\rho&=&\partial_{1}E_{1}+
(E_{1}-E_{2})\Gamma^{2}{}_{21}
+ (E_{1}-E_{3})\Gamma^{3}{}_{31}
+ H_{23}(\sigma_{3} - \sigma_{2}) \;,
\nonumber  \\
\case{1}/{3}\partial_{2}\rho&=&\partial_{2}E_{2}
+ (E_{2}-E_{1})\Gamma^{1}{}_{12}
+ (E_{2}-E_{3})\Gamma^{3}{}_{32}
+ H_{31}(\sigma_{1}-\sigma_{3}) \;,
\nonumber  \\
\case{1}/{3}\partial_{3}\rho&=&\partial_{3}E_{3}
+ (E_{3}-E_{1})\Gamma^{1}{}_{13}
+ (E_{3}-E_{2})\Gamma^{2}{}_{23}
+ H_{12}(\sigma_{2}- \sigma_{1}) \;.
\label{apb:b14}
\end{eqnarray}
to obtain the following form of (\ref{apb:b13})
\begin{eqnarray}
0 &=&
 3\sigma_{1}\partial_{1}\sigma_{1}
  -\partial_{1}(\sigma_{1}{}^{2}+\sigma_{2}{}^{2}+\sigma_{3}{}^{2})
\nonumber \\
 &&
 -\Gamma^{2}{}_{21}(\sigma_{1}- \sigma_{2})(\sigma_{3}- \sigma_{2})
 -\Gamma^{3}{}_{31}(\sigma_{1}- \sigma_{3})(\sigma_{2}- \sigma_{3})
\; .
\label{apb:b15}
\end{eqnarray}
The first term on the right of (\ref{apb:b15}) simplifies as
follows
\begin{eqnarray}
 3\sigma_{1}\partial_{1}\sigma_{1}
  -\partial_{1}(\sigma_{1}{}^{2}+\sigma_{2}{}^{2}+\sigma_{3}{}^{2})
\nonumber \\
 &=& 3\sigma_{1}\partial_{1}\sigma_{1}
  -2\sigma_{1}\partial_{1}\sigma_{1}
  -\partial_{1}(\sigma_{2}{}^{2}+\sigma_{3}{}^{2})
\nonumber \\
 &=& \sigma_{1}\partial_{1}\sigma_{1}
  -\partial_{1}(\sigma_{2}{}^{2}+\sigma_{3}{}^{2})
\nonumber \\
 &=&\case{1}/{2}\partial_{1}
     (-\sigma_{1}{}^{2}-2\sigma_{2}{}^{2}-2\sigma_{3}{}^{2})
\nonumber \\
 &=&\case{1}/{2}\partial_{1}
     [(\sigma_{2}-\sigma_{3})^{2}-2\sigma_{2}{}^{2}-2\sigma_{3}{}^{2}]
\nonumber \\
 &=&- \case{1}/{2}\partial_{1}
     (\sigma_{2}{}^{2}- 2\sigma_{2}\sigma_{3} + \sigma_{3}{}^{2})
\nonumber \\
 &=&- \case{1}/{2}\partial_{1} (\sigma_{2}{} -  \sigma_{3})^{2}
\nonumber \\
 &=&- (\sigma_{2} -  \sigma_{3})\partial_{1} (\sigma_{2} - \sigma_{3})
\; ,
\label{apb:b16}
\end{eqnarray}
so now (\ref{apb:b16}) in (\ref{apb:b15}) gives
\begin{eqnarray}
0 &=& (\sigma_{2}- \sigma_{3})\left[
  \partial_{1} (\sigma_{3} -  \sigma_{2})
 + \Gamma^{2}{}_{21}(\sigma_{1}- \sigma_{2})
 - \Gamma^{3}{}_{31}(\sigma_{1}- \sigma_{3}) \right]
\nonumber  \\
  &=& (\sigma_{2}- \sigma_{3}) H_{23}
\; ,
\label{apb:b17}
\end{eqnarray}
where in the last step we used the ``$H_{23}$'' constraint (27)
\begin{eqnarray}
H_{23} &=& \case{1}/{2} \left[ \partial_{1} (\sigma_{2} -  \sigma_{3})
 - \Gamma^{2}{}_{21}(\sigma_{1}- \sigma_{2})
 + \Gamma^{3}{}_{31}(\sigma_{1}- \sigma_{3}) \right]
\; .
\label{apb:b18}
\end{eqnarray}
Similar relations to (\ref{apb:b17}) follows for $e=2,3$ in
(\ref{apb:b3}) these are
\begin{eqnarray}
0 &=& (\sigma_{3}- \sigma_{1}) H_{31}\,,
\nonumber  \\
0 &=& (\sigma_{1}- \sigma_{2}) H_{12}
\; ,
\label{apb:b19}
\end{eqnarray}
Thus we have derived the required equations (32).

\section{The time derivative of the $H_{ab}$ constraint}
We first write the covariant form (5) of the ``$H_{ab}$'' constraint as
\begin{eqnarray}
H_{ad} &=&- h^t{}_ah^s{}_d \sigma^{b}{}_{(t}{}^{;c}\eta_{s)fbc}u^{f} \;.
   \nonumber  \\
    & =&- \case{1}/{2} h^t{}_ah^s{}_d u^{f} \left[
  \sigma^b{}_{t;c}\eta_{sfb}{}^{c} + \sigma^b{}_{s;c}\eta_{tfb}{}^{c}
                     \right ]  \; .
\label{ap:b1}
\end{eqnarray}
If we use
\begin{eqnarray}
(\sigma^{b}{}_{s;c}\dot{)} = (\dot{\sigma}^{b}{}_{s})_{;c}
                            - \sigma^{b}{}_{s;p}u^{p}{}_{;c}
+ R^{b}{}_{qpc}\sigma^{q}{}_{s}u^{p} - R^{q}{}_{spc}\sigma^{b}{}_{q}u^{p}
\; ,
\label{ap:b2}
\end{eqnarray}
the time propagation of (\ref{ap:b1}) becomes
\begin{eqnarray}
\dot{H}_{ad} = - \case{1}/{2} h^t{}_ah^s{}_d u^{f} \left[
  \eta_{sfb}{}^{c} \left\{
   (\dot{\sigma}^{b}{}_{t})_{;c} - \sigma^{b}{}_{t;p}u^{p}{}_{;c}
                   \right\}                         \right.
+ \eta_{tfb}{}^{c} \left\{
   (\dot{\sigma}^{b}{}_{s})_{;c}-\sigma^{b}{}_{s;p}u^{p}{}_{;c}
                        \right\}
        \left.             \right]
\nonumber  \\
   - \case{1}/{2} h^t{}_ah^s{}_d u^{f} \left[
  \eta_{sfb}{}^{c} \left\{
    R^{b}{}_{qpc}\sigma^{q}{}_{t}u^{p} - R^{q}{}_{tpc}\sigma^{b}{}_{q}u^{p}
                   \right\}            \right.
  + \eta_{tfb}{}^{c}\left\{
    R^{b}{}_{qpc}\sigma^{q}{}_{s}u^{p} - R^{q}{}_{spc}\sigma^{b}{}_{q}u^{p}
                   \right\}
        \left.             \right] \; .
\label{ap:b3}
\end{eqnarray}
Using the Riemann tensor $R^{s}{}_{mpc}$ expression as given in the
previous Appendix,
the two terms involving the Riemann tensor $R^{a}{}_{bcd}$
in (\ref{ap:b3}) may be written in covariant form as
\begin{eqnarray}
(RT)_{ad} &=&  - \case{1}/{2} h^t{}_ah^s{}_d u^{f} \left[
   \eta_{sfb}{}^{c} \left\{
    R^{b}{}_{qpc}\sigma^{q}{}_{t}u^{p} - R^{q}{}_{tpc}\sigma^{b}{}_{q}u^{p}
                   \right\}   \right.
 \nonumber  \\
         & &  \left.
  + \eta_{tfb}{}^{c}\left\{
    R^{b}{}_{qpc}\sigma^{q}{}_{s}u^{p} - R^{q}{}_{spc}\sigma^{b}{}_{q}u^{p}
                   \right\}
                   \right]
\nonumber  \\
 & &  \nonumber\\
 &=&   - \case{1}/{2} h^t{}_ah^s{}_d u^{f}
 \eta_{sfb}{}^{c}\left[
 \case{1}/{2}\sigma^{q}{}_{t}
     \left(g^{b}{}_{p}R_{qc} + g^{b}{}_{p}R_{qp}
                -g_{qp}R^{b}{}_{c} + g_{qc}R^{b}{}_{p} \right) \right.
\nonumber \\
    &     &
 - \case{1}/{2}\sigma^{b}{}_{q}
     \left(g^{q}{}_{p}R_{tc} + g^{q}{}_{c}R_{tp}
                -g_{tp}R^{q}{}_{c} + g_{tc}R^{q}{}_{p} \right)
\nonumber \\
       &  &
- \case{R}/{6} \sigma^{q}{}_{t}
    \left(g^{b}{}_{p}g_{qc} - g^{b}{}_{c}g_{qp}\right)
+ \case{R}/{6} \sigma^{b}{}_{q}
    \left(g^{q}{}_{p}g_{tc} - g^{q}{}_{c}g_{tp}\right)
\nonumber   \\
 & &  \nonumber\\
       & + &
 u^{i}u^{k} E^{jl}\left\{
\sigma^{q}{}_{t}\eta^{b}{}_{qij}\eta_{pckl}
+\sigma^{q}{}_{t}\left(g^{b}{}_{i}g_{qj} - g^{b}{}_{j}g_{qi}\right)
    \left(g_{pk}g_{cl} - g_{pl}g_{ck}\right)  \right.
\nonumber  \\
       &  &  \left.
-\sigma^{b}{}_{q}\eta^{q}{}_{tij}\eta_{pckl}
-\sigma^{b}{}_{q} \left(g^{q}{}_{i}g_{tj} - g^{q}{}_{j}g_{ti} \right)
    (g_{pk}g_{cl} - g_{pl}g_{ck})
               \right\}
\nonumber   \\
 & &  \nonumber\\
       &+  &
 u^{i}u^{k} H^{jl}\left\{
\sigma^{q}{}_{t} \eta^{b}{}_{qij} \left(g_{pk}g_{cl} - g_{pl}g_{ck}\right)
-\sigma^{b}{}_{q} \eta^{q}{}_{tij} \left(g_{pk}g_{cl} - g_{pl}g_{ck}\right)
  \right.
\nonumber  \\
      &   & \left.  \left.
+ \sigma^{q}{}_{t}\eta_{pckl}\left(g^{b}{}_{i}g_{qj} - g^{b}{}_{j}g_{qi}\right)
-\sigma^{b}{}_{q} \eta_{pckl}\left(g^{q}{}_{i}g_{tj} - g^{q}{}_{j}g_{ti}\right)
                   \right\}
                                \right]
\nonumber  \\
 & &  \nonumber\\
& - &
  \case{1}/{2} h^t{}_ah^s{}_d u^{f}\eta_{tfb}{}^{c}\left[
 \case{1}/{2}\sigma^{q}{}_{s}
     \left( g^{b}{}_{p}R_{qc} + g^{b}{}_{p}R_{qp}
                -g_{qp}R^{b}{}_{c} + g_{qc}R^{b}{}_{p} \right)  \right.
\nonumber \\
 &   &
 - \case{1}/{2}\sigma^{b}{}_{q}
     \left( g^{q}{}_{p}R_{sc} + g^{q}{}_{c}R_{sp}
                -g_{sp}R^{q}{}_{c} + g_{sc}R^{q}{}_{p} \right)
\nonumber \\
  &  &
-\case{R}/{6} \sigma^{q}{}_{s}
    \left(g^{b}{}_{p}g_{qc} - g^{b}{}_{c}g_{qp}\right)
+\case{R}/{6} \sigma^{b}{}_{q}
    \left(g^{q}{}_{p}g_{sc} - g^{q}{}_{c}g_{sp}\right)
\nonumber   \\
 & &  \nonumber\\
    & +  &
 u^{i}u^{k} E^{jl}\left\{
\sigma^{q}{}_{s}\eta^{b}{}_{qij}\eta_{pckl}
+\sigma^{q}{}_{s}\left(g^{b}{}_{i}g_{qj} - g^{b}{}_{j}g_{qi}\right)
    \left(g_{pk}g_{cl} - g_{pl}g_{ck}\right)  \right.
\nonumber  \\
    &  &\left.
- \sigma^{b}{}_{q}\eta^{q}{}_{sij}\eta_{pckl}
-\sigma^{b}{}_{q}\left(g^{q}{}_{i}g_{sj} - g^{q}{}_{j}g_{si}\right)
    \left(g_{pk}g_{cl} - g_{pl}g_{ck}\right)
                  \right\}
\nonumber   \\
 & &  \nonumber\\
    &+  &
 u^{i}u^{k} H^{jl}\left\{   \right.
 \sigma^{q}{}_{s} \eta^{b}{}_{qij} \left(g_{pk}g_{cl} - g_{pl}g_{ck}\right)
-\sigma^{b}{}_{q} \eta^{q}{}_{sij} \left(g_{pk}g_{cl} - g_{pl}g_{ck}\right)
\nonumber  \\
   &  & \left.  \left.
+\sigma^{q}{}_{s}\eta_{pckl}\left(g^{b}{}_{i}g_{qj} - g^{b}{}_{j}g_{qi}\right)
-\sigma^{b}{}_{q} \eta_{pckl}\left(g^{q}{}_{i}g_{sj} - g^{q}{}_{j}g_{si}\right)
                    \right\}
                                \right]
\; .\label{ap:b7a}
\end{eqnarray}
The only non-vanishing terms in (\ref{ap:b7a}) are those containing
$H^{jl}$ i.e,
\begin{eqnarray}
(RT)_{ad}&=&
- \case{1}/{2} h^t{}_ah^s{}_d u^{f}\eta_{sfb}{}^{c}u^{i}u^{k} H^{jl}
\left( \sigma^{q}{}_{t} \eta^{b}{}_{qij} g_{pk}g_{cl}
    - \sigma^{b}{}_{q} \eta^{q}{}_{tij} g_{pk}g_{cl}
\right)
\nonumber  \\
& & - \case{1}/{2} h^t{}_ah^s{}_d u^{f}\eta_{tfb}{}^{c}u^{i}u^{k} H^{jl}
\left( \sigma^{q}{}_{s} \eta^{b}{}_{qij} g_{pk}g_{cl}
    -  \sigma^{b}{}_{q} \eta^{q}{}_{sij} g_{pk}g_{cl}
\right)
\nonumber  \\
&=& -\case{1}/{2}h^t{}_ah^s{}_du^{f} u^{p}u^{i}u^{k} H^{j}{}_{c}
   \left[ \eta_{sfb}{}^{c} \left(
    g_{pk}\eta^{b}{}_{qij} \sigma^{q}{}_{t}
  - g_{pk}\eta^{q}{}_{tij} \sigma^{b}{}_{q}
                                               \right) \right.
  \nonumber  \\
& &   +   \eta_{tfb}{}^{c} \left(
    g_{pk} \eta^{b}{}_{qij} \sigma^{q}{}_{s}
  - g_{pk} \eta^{q}{}_{sij} \sigma^{b}{}_{q}
                                            \right)
     \left.             \right]
  \nonumber  \\
&=& \case{1}/{2} h^t{}_ah^s{}_d u^{f}u^{i} H^{j}{}_{c}
    \left[
      \eta_{sfb}{}^{c}\left(
   \eta^{b}{}_{qij} \sigma^{q}{}_{t}
  - \eta^{q}{}_{tij} \sigma^{b}{}_{q}
                                               \right) \right.
  \nonumber  \\
& &     +   \eta_{tfb}{}^{c}u^{f} \left(
     \eta^{b}{}_{qij} \sigma^{q}{}_{s}
  -  \eta^{q}{}_{sij} \sigma^{b}{}_{q}
                                            \right)
     \left.             \right]
\; .\label{ap:b7}
\end{eqnarray}
Equation (\ref{ap:b7}) represent the covariant form of the Riemann term.
The tetrad form of (\ref{ap:b7}) for {\bf diagonal} elements $a = d$
becomes
\begin{eqnarray}
(RT)_{aa} &=& \eta_{a0b}{}^{c}\eta^{b}{}_{a0j} H^{j}{}_{c}
     ( \sigma_{a}-\sigma_{b} )
\; .\label{ap:b8}
\end{eqnarray}
and hence
\begin{eqnarray}
(RT)_{11}&=& H_{3}(\sigma_{1} -\sigma_{2}) + H_{2}(\sigma_{1} -\sigma_{3})
\; , \nonumber \\
(RT)_{22}&=& H_{1}(\sigma_{2} -\sigma_{3}) + H_{3}(\sigma_{2} -\sigma_{1})
\; , \nonumber \\
(RT)_{33}&=& H_{2}(\sigma_{3} -\sigma_{1}) + H_{1}(\sigma_{3} -\sigma_{2})
\; .
\label{ap:b9}
\end{eqnarray}
The tetrad form of the time propagation equation  (\ref{ap:b3}) for $a=d$
now becomes
\begin{eqnarray}
\dot{H}_{aa} &=&- \eta_{a0b}{}^{c}\left[
  \partial_{c}\dot{\sigma}^{b}{}_{a} - \theta_{c}\partial\sigma^{b}{}_{a}
  + \Gamma^{b}{}_{ca} \left\{ (\dot{\sigma}_{a} -\dot{\sigma}_{b})
  \right\}
    -\theta_{c}(\sigma_{a}-\sigma_{b}) \right]
\nonumber \\
& & + (RT)_{aa} \;\;\;\;\; \mbox{(no sum over)}\; a
\; .
\label{ap:b10}
\end{eqnarray}
For $a=1$ (\ref{ap:b10}) becomes
\begin{eqnarray}
\dot{H}_{11} &=&
   \Gamma^{3}{}_{21} \left[ (\dot{\sigma}_{1} -\dot{\sigma}_{3})
    -\theta_{2}(\sigma_{1}-\sigma_{3}) \right]
   - \Gamma^{2}{}_{31} \left[ (\dot{\sigma}_{1} -\dot{\sigma}_{2})
    -\theta_{3}(\sigma_{1}-\sigma_{2}) \right]
    +(RT)_{11}
\; .
\nonumber \\
\label{ap:b11}
\end{eqnarray}
On using $\dot{\sigma}_{\mu}$ in (20) and constraint
(26) we get
\begin{eqnarray}
\dot{H}_{11} &=& -\theta H_{11} + I_{11} + (RT)_{11}
\; .
\label{ap:b12}
\end{eqnarray}
If we compare with the propagation equation for $\dot{H}_{11}$ in
(21) i.e.,
\begin{eqnarray}
\dot{H}_{11} &=& -\theta H_{11} + I_{11}
+ 3\sigma_{1}H_{11} - (\sigma_{1}H_{11} + \sigma_{2}H_{22} + \sigma_{3}H_{33})
\; ,
\label{ap:b13}
\end{eqnarray}
we get
\begin{eqnarray}
3\sigma_{1}H_{11} - (\sigma_{1}H_{11} + \sigma_{2}H_{22} + \sigma_{3}H_{33})
\nonumber \\
= (RT)_{11}= H_{3}(\sigma_{1} -\sigma_{2}) + H_{2}(\sigma_{1} -\sigma_{3})
\; .
\label{ap:b14}
\end{eqnarray}
We note however that
\begin{eqnarray}
3\sigma_{1}H_{11} - (\sigma_{1}H_{11} + \sigma_{2}H_{22} + \sigma_{3}H_{33})
&=& 2\sigma_{1}H_{11} - \sigma_{2}H_{22} - \sigma_{3}H_{33}
\nonumber \\
&=& [\sigma_{1}H_{11} - \sigma_{2}H_{22}]
   +[\sigma_{1}H_{11} - \sigma_{3}H_{33}]
\nonumber \\
&=& [\sigma_{1}H_{11} +(\sigma_{1}+\sigma_{3}) H_{22}]
  + [\sigma_{1}H_{11} +(\sigma_{1}+\sigma_{2})H_{33}]
\nonumber \\
&=& [\sigma_{1}(H_{11}+H_{22})  + \sigma_{3} H_{22}]
 +  [\sigma_{1}(H_{11}+H_{33}) + \sigma_{2}H_{33}]
\nonumber \\
&=& -\sigma_{1}H_{33} + \sigma_{3} H_{22}
  -\sigma_{1}H_{22} + \sigma_{2}H_{33}
\nonumber \\
&=& H_{33}(\sigma_{2}-\sigma_{1})
+  H_{22}(\sigma_{3} -\sigma_{1})
\nonumber \\
&=& - (RT)_{11}
\; .
\label{ap:b15}
\end{eqnarray}
Thus (\ref{ap:b14}) together with (\ref{ap:b15}) gives
\begin{eqnarray}
0 &=& 3\sigma_{1}H_{11} - (\sigma_{1}H_{11} + \sigma_{2}H_{22} +
\sigma_{3}H_{33})
\; ,\label{ap:b16} \\
0&=& (RT)_{11} = H_{3}(\sigma_{1} -\sigma_{2}) + H_{2}(\sigma_{1} -\sigma_{3})
\; .
\label{ap:b16a}
\end{eqnarray}
Similar results hold for $a=d=2$ and $a=d=3$ respectively
\begin{eqnarray}
0&=& 3\sigma_{2}H_{22} - (\sigma_{1}H_{11} + \sigma_{2}H_{22} +
\sigma_{3}H_{33})
\; ,\label{ap:b17} \\
0&=& (RT)_{22}=H_{1}(\sigma_{2} -\sigma_{3}) + H_{3}(\sigma_{2} -\sigma_{1})
\; .\label{ap:b17a}
\end{eqnarray}
\begin{eqnarray}
0&=& 3\sigma_{3}H_{33} - (\sigma_{1}H_{11} + \sigma_{2}H_{22} +
\sigma_{3}H_{33})
\; ,\label{ap:b18} \\
0&=& (RT)_{33}=H_{2}(\sigma_{3} -\sigma_{1}) + H_{1}(\sigma_{3} -\sigma_{2})
\; .\label{ap:b18a}
\end{eqnarray}

If we write (\ref{ap:b16a}),(\ref{ap:b17a}),(\ref{ap:b18a}) together
we get
\begin{eqnarray}
0&=& (RT)_{11} = H_{3}(\sigma_{1} -\sigma_{2}) + H_{2}(\sigma_{1} -\sigma_{3})
\; ,
\label{ap:b16b}  \\
0&=& (RT)_{22}=H_{1}(\sigma_{2} -\sigma_{3}) + H_{3}(\sigma_{2} -\sigma_{1})
\; ,
\label{ap:b17b} \\
0&=& (RT)_{33}=H_{2}(\sigma_{3} -\sigma_{1}) + H_{1}(\sigma_{3} -\sigma_{2})
\; ,
\label{ap:b18b}
\end{eqnarray}
which are consistent with each other.


Similarly if we write equations (\ref{ap:b16}),(\ref{ap:b17}),(\ref{ap:b18})
which are equivalent to (\ref{ap:b16a}),(\ref{ap:b17a}),(\ref{ap:b18a})
we get
\begin{eqnarray}
0 &=& 3\sigma_{1}H_{11} - (\sigma_{1}H_{11} + \sigma_{2}H_{22} +
\sigma_{3}H_{33})
\; ,\label{ap:b16c} \\
0&=& 3\sigma_{2}H_{22} - (\sigma_{1}H_{11} + \sigma_{2}H_{22} +
\sigma_{3}H_{33})
\; ,\label{ap:b17c} \\
0&=& 3\sigma_{3}H_{33} - (\sigma_{1}H_{11} + \sigma_{2}H_{22} +
\sigma_{3}H_{33})
\; ,\label{ap:b18c} \\
\end{eqnarray}
from which  it follows that
\begin{eqnarray}
\sigma_{1}H_{11} =\sigma_{2}H_{22}=\sigma_{3}H_{33}
\; .\label{ap:b19}
\end{eqnarray}
Equation (\ref{ap:b19}), which is the required equation (37),
is a condition  arising from consistency requirement on the
``$H_{ad}$'' for $a=d$.

Similarly  for {\bf non-diagonal} elements $a\neq b$ equation
(\ref{ap:b7})
\begin{equation}
(RT)_{23} = \case{3}/{2}\sigma_{1}H_{23} \; ,
\nonumber \\
(RT)_{31} = \case{3}/{2}\sigma_{2}H_{31} \; ,
\nonumber \\
(RT)_{12} = \case{3}/{2}\sigma_{3}H_{12} \; ,
\label{ap:b21}
\end{equation}
and the time propagation equation (\ref{ap:b3})
yields
\begin{eqnarray}
\sigma_{1}H_{23}= 0\;,\;\;\; (a=2,d=3)\; ;
\nonumber \\
\sigma_{2}H_{13}= 0\;,\;\;\; (a=1,d=3)\; ;
\nonumber \\
\sigma_{3}H_{12}= 0\;,\;\;\; (a=1,d=2)\; .
\label{ap:b22}
\end{eqnarray}
which is the required equation (34).

\begin{center}
-------------------------
\end{center}

\end{document}